\documentclass[12pt]{article}

\pdfoutput=1

\usepackage[bulletsep]{collref}
\usepackage{amssymb,graphicx}
\usepackage[intlimits]{amsmath}
\usepackage{bbm}
\usepackage[small]{subfigure}

\usepackage{MnSymbol}

\makeatletter \@addtoreset{equation}{section} \makeatother

\makeatletter
\let\old@startsection=\@startsection
\let\oldl@section=\l@section
\renewcommand{\@startsection}[6]{\old@startsection{#1}{#2}{#3}{#4}{#5}{#6\mathversion{bold}}}
\renewcommand{\l@section}[2]{\oldl@section{\mathversion{bold}#1}{#2}}
\makeatother

\makeatletter
\let\old@makecaption=\@makecaption
\def\@makecaption{\small\old@makecaption}
\makeatother

\begin{document}

\newcommand{\be}{\begin{equation}}\newcommand{\ee}{\end{equation}}
\newcommand{\bea}{\begin{eqnarray}} \newcommand{\eea}{\end{eqnarray}}
\def\p{\partial}
\def\pa{\partial}
\def\ov{\over }
\def\a{\alpha }
\def\g{\gamma}
\def\s{\sigma }
\def\td{\tilde }
\def\vp{\varphi}
\def\gd{\nu }
\def \ha {{1 \over 2}}

\def\Xint#1{\mathchoice
{\XXint\displaystyle\textstyle{#1}} 
{\XXint\textstyle\scriptstyle{#1}} 
{\XXint\scriptstyle\scriptscriptstyle{#1}} 
{\XXint\scriptscriptstyle\scriptscriptstyle{#1}} 
\!\int}
\def\XXint#1#2#3{{\setbox0=\hbox{$#1{#2#3}{\int}$ }
\vcenter{\hbox{$#2#3$ }}\kern-.5\wd0}}
\def\ddashint{\Xint=}
\def\dashint{\Xint-}

\newcommand\cev[1]{\overleftarrow{#1}} 

\begin{flushright}\footnotesize
%\texttt{ITEP-TH-18/11}\\
\texttt{NORDITA-2012-57} \\
\texttt{UUITP-20/12}
\vspace{0.6cm}
\end{flushright}

\renewcommand{\thefootnote}{\fnsymbol{footnote}}
\setcounter{footnote}{0}

\begin{center}
{\Large\textbf{\mathversion{bold} Large $N$ Limit of $\mathcal{N}=2$ $SU(N)$ Gauge 
\\ Theories  from Localization}
\par}

\vspace{0.8cm}

\textrm{J.G.~Russo$^{1,2}$\footnote{On leave of absence from Universitat de Barcelona and Institute of Cosmos Sciences, Barcelona, Spain.} and
K.~Zarembo$^{3,4}$\footnote{Also at ITEP, Moscow, Russia.}}
\vspace{4mm}

\textit{${}^1$ Perimeter Institute for Theoretical Physics,\\
Waterloo, Ontario, N2L 2Y5, Canada}\\
\textit{${}^2$ Instituci\'o Catalana de Recerca i Estudis Avan\c cats (ICREA), \\
Pg. Lluis Companys, 23, 08010 Barcelona, Spain}\\
\textit{${}^3$Nordita,
Roslagstullsbacken 23, SE-106 91 Stockholm, Sweden}\\
\textit{${}^4$Department of Physics and Astronomy, Uppsala University\\
SE-751 08 Uppsala, Sweden}\\
\vspace{0.2cm}
\texttt{jorge.russo@icrea.cat, zarembo@nordita.org}
%\vspace{3mm}

\vspace{3mm}

%%%%%%%%

\par\vspace{1cm}

\textbf{Abstract} \vspace{3mm}

\begin{minipage}{13cm}
We study $\mathcal{N}=2$ Yang-Mills theory on $S^4$ in the large-$N$ limit. We find that on a large sphere Wilson loops obey a perimeter law and that the free energy grows quadratically with the radius of the sphere. We also comment on the large-$N$ limit of the $\mathcal{N}=2^*$ theory, and on the free energy in $\mathcal{N}=2$ and $\mathcal{N}=4$ superconformal theories.
\end{minipage}

\end{center}

\vspace{0.5cm}

%%%%%%%%%%%%%%%%%%%%%%%%%%%%%%%%%%%%%%%%%%%%%%%%%%%%%%%%%%%%%%%%%%%%%%%%%%%%%%%%
\clearpage

\tableofcontents

\setcounter{page}{1}
\renewcommand{\thefootnote}{\arabic{footnote}}
\setcounter{footnote}{0}

\section{Introduction}

In supersymmetric field  theories a certain class of non-trivial dynamical quantities can be computed exactly. An example, that we will further explore here, is the partition function of an $\mathcal{N}=2$ supersymmetric gauge theory  on $S^4$. By using localization, the $S^4$ partition function can be reduced to a finite-dimensional matrix integral  \cite{Pestun:2007rz}. Observables, that can be computed by this method, include the free energy and expectation values of circular Wilson loops.

 One can envision a number of applications of this remarkable result. The 't~Hooft's large-$N$ limit of the matrix integral \cite{Rey:2010ry,Passerini:2011fe,Bourgine:2011ie,Fraser:2011qa,Russo:2012kj} may give us a glimpse on the putative string duals of $\mathcal{N}=2$ theories. In particular, for the $\mathcal{N}=4$ super-Yang-Mills theory (SYM), localization confirms earlier conjectures on the circular Wilson loop \cite{Erickson:2000af,Drukker:2000rr}, which were instrumental in verifying the AdS/CFT correspondence. Or one can address generic questions about quantum field theory, such as the asymptotic behavior of perturbative series \cite{Russo:2012kj} in the setting where exact non-perturbative results are available. 

We will study the large-$N$ limit of the Pestun's matrix model, focusing on the simplest example of asymptotically free gauge theory, pure $\mathcal{N}=2$ SYM. 
When the radius of the four-sphere becomes infinite we will be able to compare our results to the large-$N$ limit of Seiberg-Witten theory \cite{Seiberg:1994rs}, which was analyzed in \cite{Douglas:1995nw,Ferrari:2001mg}. 
We will also comment on the $\mathcal{N}=2^*$ theory (a massive deformation of $\mathcal{N}=4$), which was studied in \cite{Russo:2012kj}, and on the superconformal theories considered in  \cite{Rey:2010ry,Passerini:2011fe,Bourgine:2011ie,Fraser:2011qa}. The motivation here is the AdS/CFT duality, and  we will fill the gap in the literature by computing the finite part of the holographic free energy in $\mathcal{N}=4$ SYM.

What do we expect to find, when we study an asymptotically free QFT on $S^4$?
In absence of other parameters in the problem, the dynamics is controlled by the running coupling renormalized at the inverse size of the sphere ($R^{-1}$). When the radius is small, perturbation theory should  work and yield an expansion of observables in  inverse powers of  $\ln R^{-1}$. For a large sphere, the theory enters a non-perturbative regime, and the behavior of observables becomes a non-trivial dynamical question. We know that $\mathcal{N}=2$ SYM is not confining \cite{Seiberg:1994rs}, and may thus expect that Wilson loops will obey a perimeter  law, rather than  an area law. As for the free energy, it cannot be extensive (cannot scale as $R^4$), because the vacuum energy is strictly zero in any supersymmetric theory. Since the perturbative free energy quadratically diverges in $\mathcal{N}=2$ SYM (see footnote 1), we may expect to find a $\Lambda^2 R^2$ behavior, where the UV cutoff is replaced by the dynamically generated mass scale $\Lambda $. The analysis of the matrix model will confirm these expectations.

\section{$\mathcal{N}=2$ SYM}

\subsection{Partition function}

The partition function of any $\mathcal{N}=2$ supersymmertic gauge theory on $S^4$ is calculable by localization methods \cite{Pestun:2007rz}. The end result of this calculation is a finite-dimensional integral over the  moduli space of Coulomb vacua, parameterized by the eigenvalues of the scalar field $\Phi $ from the vector multiplet: $\Phi =\mathop{\mathrm{diag}}(a_1,\ldots a_N)$, $\mathop{\mathrm{tr}}\Phi =0$. We will mostly concentrate on  the simplest example, pure $\mathcal{N}=2$  Yang-Mills theory without matter multiplets. Its partition function is given by the eigenvalue integral
\begin{equation}\label{Pestunmatrix}
 Z= \int d^{N-1}a \,\prod_{i<j}^{}\left[\left(a_i-a_j\right)^2H^2(a_i-a_j)\right]\,{\rm e}\,^{-\frac { 8\pi^2 N} {\lambda }\,\sum\limits_{i} a_i ^2 }\left|{\cal Z }_\text{inst}(a;g^2)\right|^2.
\end{equation}
The key point in the localization method is the one-loop exactness of the effective action for the eigenvalues. It thus suffices to compute one-loop determinants and to sum over instantons.

The one-loop factor was calculated in  \cite{Pestun:2007rz} and is expressed in terms of the function $H(x)$, defined as an infinite product over spherical harmonics:
\be\label{Hdef}
 H(x)\equiv \prod_{n=1}^\infty 
\left(1+{x^2\over n^2}\right)^n \,{\rm e}\,^{-{x^2\over n}}.
\ee
The coupling constant in the partition function is actually the running coupling renormalized at the scale set by the radius $R$ of the sphere, the only parameter in the problem. Of course, the original path integral before localization depends on the bare coupling $\lambda _0$, but this  combines with the one-loop UV logarithms to produce the renormalized running coupling $\lambda $.
Indeed,  the product over spherical harmonics appears  in   the calculation of \cite{Pestun:2007rz} without the suppression factor $\exp(-x^2/n)$  and thus logarithmically diverges. The infinite contribution, cut off at some $n_0=\Lambda _0R$ is absorbed by renormalizing the bare coupling: 
\be
\frac{4\pi^2}{\lambda_0} -\ln\left(e^\gamma \Lambda _0R\right)  \equiv  \frac{4\pi ^2}{\lambda }\,,
\ee
as can be easily seen from (\ref{Pestunmatrix}) ($\gamma $ is the Euler constant). A more mathematically precise way to   get rid of  divergences, used in \cite{Pestun:2007rz} and also discussed later in section~\ref{2star}, is to consider $\mathcal{N}=2^*$ theory as a UV completion of pure $\mathcal{N}=2$ SYM. The mass of the adjoint hypermultiplet then plays the  r\^ole of the UV cutoff $\Lambda _0$. 

The running coupling can be expressed through the radius of the sphere and the dynamically generated finite scale $\Lambda =\Lambda _0\exp(-4\pi ^2/\lambda _0+\gamma )$:
\begin{equation}
 \frac{4\pi ^2}{\lambda }=-\ln\left(\Lambda R\right).
\end{equation}
The running coupling becomes negative at $R>\Lambda ^{-1}$. What makes the integral in (\ref{Pestunmatrix}) convergent is the rapid decrease at infinity of the one-loop function $H(x)$.

\medskip

Although it is commonly assumed that instantons are unimportant at large-$N$, this is guaranteed to be true only at weak coupling. In principle the entropy (the volume of the instanton moduli space) may overcome the exponential suppression by the instanton action. 
Then the instanton weight becomes exponentially large, and the system undergoes a transition to a new phase. Whether such transition happens or not is a dynamical question \cite{Gross:1994mr}. Such an instanton-induced phase transition does not occur in the ${\mathcal N}=2$ superconformal theory \cite{Passerini:2011fe}. 
This dynamical question is more intricate in the ${\mathcal N}=2$ SYM case,
where the coupling runs and it is not 
clear a priori which is the relevant scale for $\lambda$ appearing in the instanton weight. As explained below, the scale at which the coupling is renormalized, can get dynamically shifted from $R^{-1}\to aR^{-1}$. The new scale effectively represents the inverse of the instanton size, and  a relevant question is whether the partition function is dominated by small or large instantons.
These issues will be addressed in section 5, where the instanton factor $\left|{\cal Z }_\text{inst}\right|^2$ in the large $N$ limit  is discussed.
Meanwhile, we will first set ${\cal Z }_\text{inst}$ to $1$, and  in section 5 we will argue that indeed it is not relevant for the large $N$ dynamics by explicitly computing the one-instanton contribution.

\medskip

Supersymmetric observables map, upon localization, onto expectation values in the matrix model. For example, the Wilson loop in the fundamental representation on the big circle of $S^4$ maps to the trace of the matrix exponential:
\begin{equation}
\label{Wil}
 W\left(C_{\rm circle}\right)=\left\langle \sum_{i=1}^{N}
 \,{\rm e}\,^{2\pi a_i}
 \right\rangle.
\end{equation}
Our goal is to solve the matrix model in the large-$N$ limit and to compute the free energy and the Wilson loop vacuum expectation value (VEV). 

\subsection{Toy model: $SU(2)$}

Before proceeding with the large-$N$ calculation, it is instructive to consider the $SU(2)$ partition function, which is given by a simple one-dimensional integral:
\begin{equation}\label{SU2}
 Z_{SU(2)}^{k=0}=\int_{-\infty }^{+\infty }da\,
 a^2H^2(2a)\left(\Lambda R\right)^{8a^2},
\end{equation}
which we will compute in the saddle-point approximation. As we will explain, the saddle-point approximation is not arbitrarily accurate for $SU(2)$, but numerically works pretty well.  A more serious problem is that one cannot really ignore instantons at small $N$, so  this  is just a toy model that illustrates the effect of the one-loop determinant. It gives a correct qualitative picture of what happens in the $SU(N)$ theory at large $N$. 

The saddle point of the integral (\ref{SU2}) is determined by a minimum of the effective potential:
\begin{equation}
 S_{\rm eff}(a)=-8a^2\ln\left( \Lambda R\right)-2\ln H(2a)\ .
\end{equation}
Here we ignore a term $-\ln a^2$ coming from the Vandermonde determinant, which can be viewed as a quantum prefactor.
It is not significant for the large $\Lambda R$ behavior.

\medskip

\begin{figure}[t]
\begin{center}
 \centerline{\includegraphics[width=10cm]{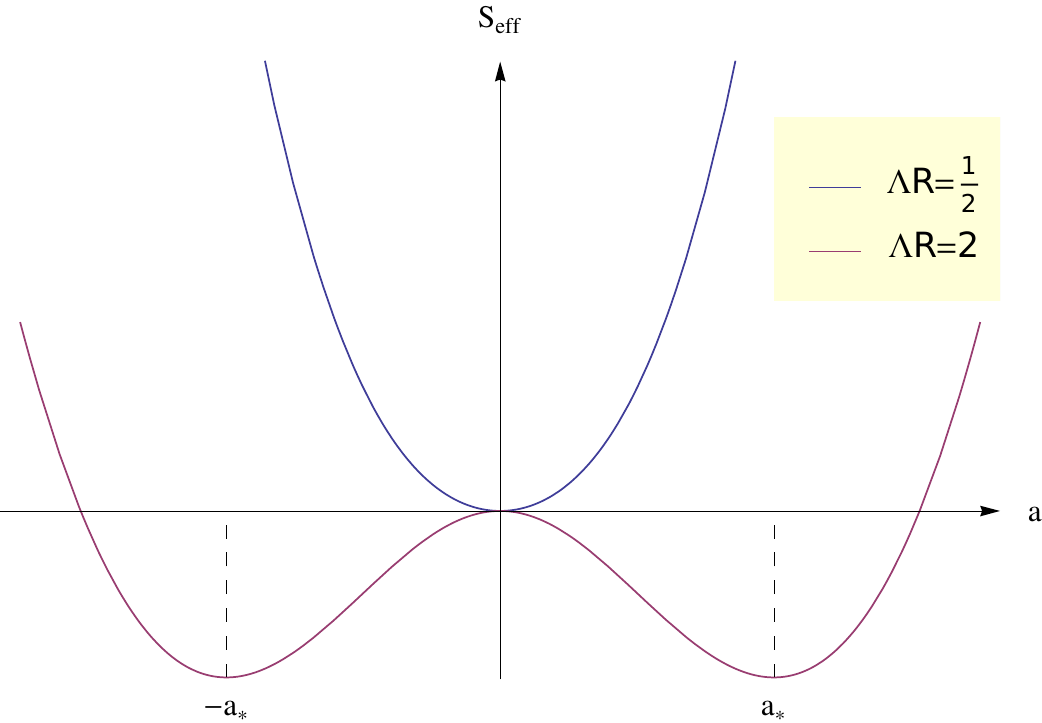}}
\caption{\label{seff}\small The effective potential of the $SU(2)$ model for $\Lambda R=1/2$ and $\Lambda R=2$.}
\end{center}
\end{figure}

The function $H(x)$ has the following large and small $x$ asymptotics:
\begin{equation}
 \ln H(x)=
\begin{cases}
  -\frac{\zeta (3)}{2}\,x^4+O\left(x^6\right)& {\rm at}~x\rightarrow 0
  \\
 -\,x^2\ln |x|\,{\rm e}\,^{\gamma -\frac{1}{2}} +O\left(\ln x\right)& {\rm at}~x\rightarrow \infty 
\end{cases}
\end{equation}
The coupling constant here is renormalized at the inverse radius of the sphere $R^{-1}$, but only at small $a$. When $a$ is large, $2\ln H(2a)$ behaves as $-8a^2\ln a$ and a new scale $aR^{-1}$ emerges. The effective potential thus has the Coleman-Weinberg shape \cite{Coleman:1973jx},  illustrated in
fig.~\ref{seff}.
The potential has a minimum at zero for  $R\Lambda <1$ (which is slightly shifted off zero if the Vandermonde determinant is included). 
However, if $R>\Lambda ^{-1}$, this point becomes a maximum, and the saddle point shifts to a solution of the  equation:
\begin{equation}
 4a_*\ln \Lambda R=K(2a_*),
\end{equation}
where 
\begin{equation}
 K(x)=-\frac{H'(x)}{H}=2x\sum_{n=1}^{\infty }\left(
 \frac{1}{n}-\frac{n}{n^2+x^2}
 \right)=x\left(\psi \left(1+ix\right)+\psi \left(1-ix\right)-2\psi (1)\right)\ ,
\end{equation}
and $\psi(1)=-\gamma$.
The saddle-point equation is essentially the condition $g_{\rm eff}(a_*R^{-1})=\infty $. 
{}For large radius, the saddle-point position $a_*$ thus grows approximately linearly with $R$, $2a_*\approx \,{\rm e}\,^{-\gamma }\Lambda R $
(the saddle-point approximation applies when the second derivative of the potential is large; in the present case,  $S''_{\rm eff}(a_*)=16+O( R^{-2})$).

\medskip

The VEV of the circular Wilson in the saddle-point approximation is given by $W(C_{\rm circle})=\cosh 2\pi a_*$. Since $a_*=0$ at $R<\Lambda ^{-1}$, the Wilson loop stays constant at small $R$, within the semiclassical approximation. In practice, the exact VEV slowly grows with $R$ until $R$ reaches $R_c\sim \Lambda ^{-1}$, after which $a_*\neq 0$ and the Wilson loop starts growing very fast. Asymptotically, at very large $R$,
\begin{equation}
 \ln W(C_{\rm circle})\simeq \pi \,{\rm e}\,^{-\gamma }\,    \Lambda R,\qquad F=S(a_*)\simeq -\,{\rm e}\,^{-2\gamma }\, \Lambda ^2R^2.
\end{equation}
The Wilson loop thus satisfies the perimeter law, and the free energy is quadratic in the radius of the sphere.
We will find the same behavior in the $SU(N)$ SYM at large $N$. 

\smallskip

The perimeter law means that the mass of a heavy probe, associated with the Wilson loop, acquires a finite negative renormalization. The quadratic dependence of the free energy on the radius can be interpreted as a finite renormalization of the Newton's constant. In fact, a short-distance, quadratically divergent contribution to the Newton's constant does not vanish in the pure $\mathcal{N}=2$ theory\footnote{This can be seen by setting $N_{\frac{1}{2}}=2N_1$ in eqs. (11), (15) and (18) in \cite{Burgess:1999vb}.}. The quadratic dependence of the free energy on the dynamically generated scale can be viewed as a finite remnant of this quadratic UV divergence. 

\subsection{Saddle-point equation}

At large $N$, the eigenvalue distribution is characterized by a continuous density $\rho \left(x\right)$, defined on a interval $(-\mu ,\mu )$. 
The integral (\ref{Pestunmatrix}) is dominated by an eigenvalue distribution  determined by
the saddle-point equation 
\be
\strokedint_{-\mu}^\mu dy \rho(y) \left(\frac{1}{x-y} -K(x-y)\right)
= \frac{8\pi^2}{\lambda}\,x
\equiv  -2x\ln\left( \Lambda R\right) .
\label{uno}
\ee
The eigenvalue density is subject to the normalization condition
\be
\int_{-\mu}^\mu dy\ \rho(y) =1.
\ee
We  now study the solutions to (\ref{uno}) in different regimes.

\subsubsection*{Small radius}

When $\Lambda R\ll 1$, one has $0<\lambda\ll 1$. This is the perturbative regime.
In this region, the linear force represented by $8\pi^2 x/\lambda $ in (\ref{uno}) pushes all eigenvalues near  
$x=0$. As a result, 
$\mu\ll 1$ and we can use the expansion of $K(x)$ in powers of $x$,
\be
\label{kara}
K(x)=  -2\sum_{k=1}^{ \infty} (-1)^k \zeta(2k+1) x^{2k+1}\ ,
\ee
The solution of the eigenvalue density  equation (\ref{uno}) is then of the form
\be
\rho(x) =\left( \sum_{k=0}^{k_{\rm max}} a_k x^{2k} \right) \sqrt{\mu^2-x^2}
\label{ansrho}
\ee
where $k_{\rm max}$ is determined by  the perturbative order that one wishes
to compute. Then (\ref{ansrho}) is a solution to  (\ref{uno}) with
 $K(x)$ defined by the sum (\ref{kara}) truncated at the same  $k_{\rm max}$.

Substituting the ansatz (\ref{ansrho}) in (\ref{uno}) we find an equation of the form
$\sum_{k=0}^{k_{\rm max}} A_k x^{2k+1}=0 $. This gives $k_{\rm max}+1$ conditions for
 $k_{\rm max}+1$ unknowns, $a_0, a_1, ...,a_{k_{\rm max}}$.
Then $\mu$ is found by the normalization condition.

Let us explicitly work out the case $k_{\rm max}=2$.
Using the formulas,
\be
\int_{-\mu}^\mu dy\ y^{2n} \sqrt{\mu^2-y^2}= \frac{\sqrt{\pi}\ \Gamma(n+{1\over 2})\mu ^{2n+2}}{2(n+1)!}\ ,
\ee
\be
\strokedint_{-\mu}^\mu dy \ \rho(y)\ \frac{1}{ (x-y)}=
x^3 \left(\pi  a_1-\frac{1}{2} \pi  a_2 \mu ^2\right)+x \left(\pi  a_0-\frac{1}{2} \pi  a_1 \mu ^2-\frac{1}{8} \pi  a_2
   \mu ^4\right)\ ,
\ee
 one gets a linear system of equations
for $a_0,\ a_1,\ a_2$.
Substituting the solution in the normalization condition leads to a polynomial equation
for $\mu $ which can be solved in a power series in $\lambda$,
\be
\label{poi}
\mu^2=\frac{\lambda }{4 \pi ^2}-\frac{3 \lambda ^3 \zeta (3)}{128 \pi ^6}
+\frac{25 \lambda ^4 \zeta (5)}{2048 \pi ^8}+O\left(\lambda ^5\right)\ .
\ee
The expectation value of the circular Wilson loop (\ref{Wil}) is then
computed from the formula
\be
W(C_{\rm circle}) = \int_{-\mu}^\mu dx \ \rho(x)\ \,{\rm e}\,^{2\pi x}\ .
\label{wils}
\ee
This gives  a linear combination
of $I_1(2\pi\mu)$ and $I_2(2\pi\mu)$ Bessel functions. Expanding in powers of $\lambda$, we obtain the perturbative series
\be
\label{gwil}
W =
1+\frac{\lambda }{8}+\frac{\lambda ^2}{192}+\lambda ^3 \left(\frac{1}{9216} - 
\frac{5 \zeta (3)}{512\pi^4}\right)
+\lambda ^4 \left(\frac{1}{737280}-\frac{7 \zeta (3)}{8192 \pi ^4}+\frac{35 \zeta (5)}{8192 \pi ^6}\right)+O\left(\lambda ^5\right)\ .
\ee
This may be compared with explicit evaluation of Feynman diagrams, except that the matrix-model calculation is enormously easier and in principle can be pushed to an arbitrary order in  $\lambda $. It is interesting to notice that the expansion can be organized differently, without expanding the Bessel functions in $\lambda $. 
This would give
\bea
W &=&
\frac{2 I_1(2 \pi  \mu )}{\sqrt{\lambda }}-
\frac{3 \lambda}{8 \pi ^6}
 \left(\pi ^2 \zeta (3)-5 \zeta (5)\right) I_2(2 \pi 
   \mu )+\frac{3 \lambda ^{3/2}}{32 \pi
   ^6} \left(2 \pi ^2 \zeta (3)-5 \zeta (5)\right) I_1(2 \pi  \mu )
\nonumber\\
&+&\frac{15 \lambda ^2 \zeta (5)}{32 \pi ^6}\, I_2(2 \pi  \mu )+O\left(\lambda ^{4}\right)\ .
\eea
Considering the leading term  in $ \mu $, $\mu\approx \sqrt{\lambda}/(2\pi)$, the first term in this expansion is the Wilson loop VEV in the $\mathcal{N}=4$ SYM, which is known to resum an infinite series of Feynman diagrams without internal vertices \cite{Erickson:2000af,Drukker:2000rr}. The $\mathcal{N}=4$ part in (\ref{gwil}) can be easily recognized as those terms that have no $\zeta $ function coefficients.
This part of the series has an infinite radius of convergence.
\smallskip

It would be very interesting to determine the asympotic behavior of the full series   in  (\ref{gwil}).
 It is  believed that in gauge theories  the number of  planar graphs of $n$th order  grows like a power of $n$, as opposed to the $n!$ total  number of Feynman diagrams \cite{Makeenko:2002uj}. This indicates that, if  the coefficient of the  $\lambda ^n$ term grows as $n!$, as occurs in the $SU(2)$ cases studied in \cite{Russo:2012kj}, 
 it would be  be due to  UV renormalons.\footnote{ 
Infrared renormalons arise from the small momentum part of the loop integrals in a certain subset of $n$-loop Feyman diagrams. They should not appear in the present context because the radius of the sphere provides an IR cutoff; integrals over momenta are replaced by  sums over
spin 0, 1/2, and 1 spherical harmonics of $S^4$.} In this case the planar series  (\ref{gwil}) would be Borel summable, since UV renormalons 
should not produce poles on the real positive axes of the Borel plane (this expectation is consistent with the alternating sign  character of the series).

\subsubsection*{Large radius}

\begin{figure}[h!]
 \centering
 \includegraphics[width=.45\textwidth]{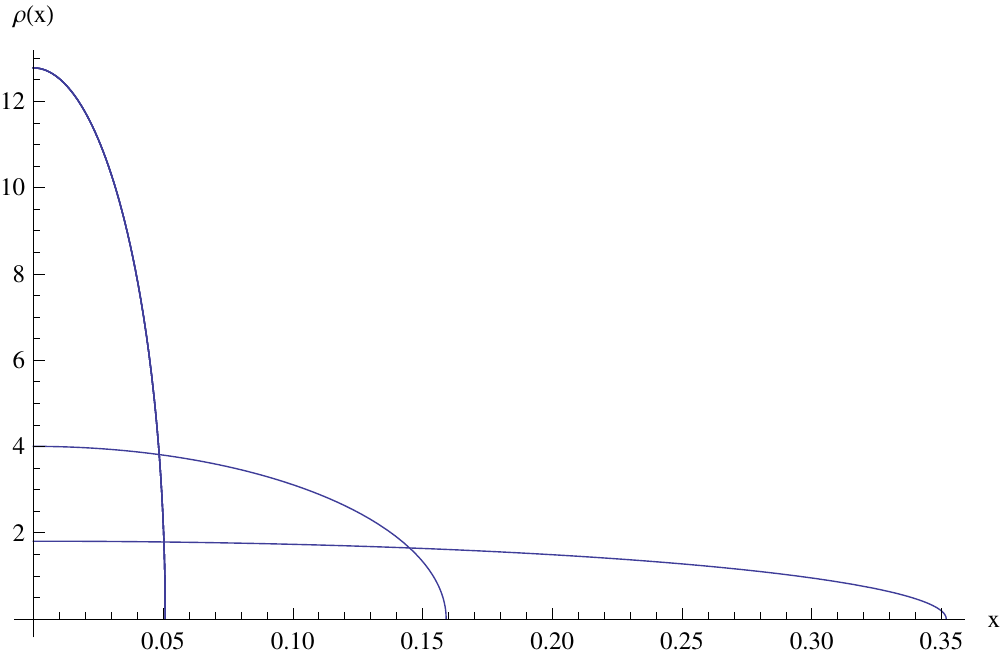}
 \includegraphics[width=.45\textwidth]{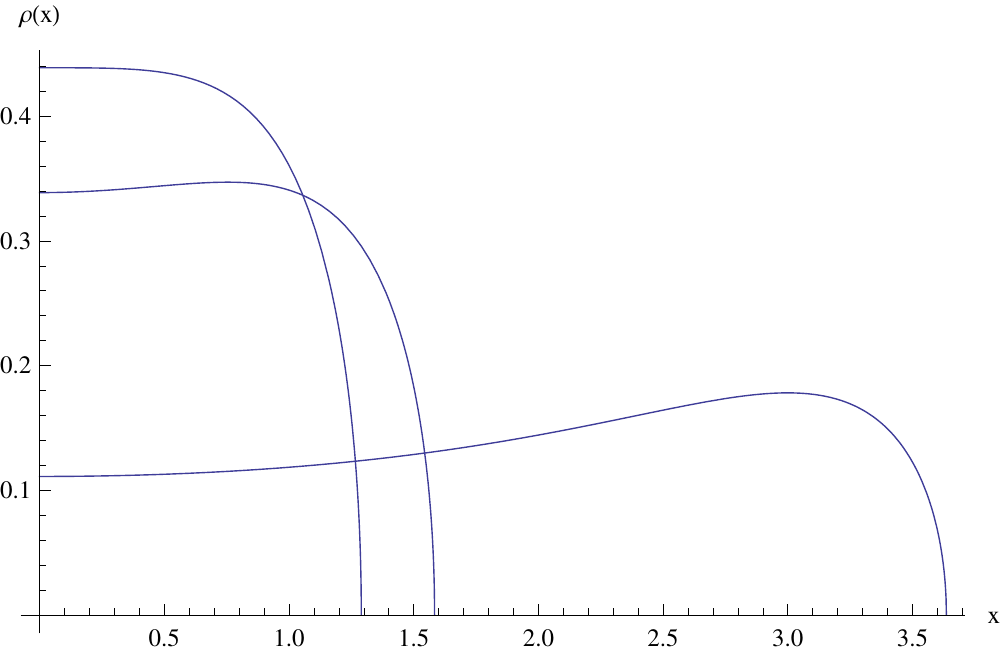} 
 \caption{ \label{densityprofiles}
a) Cases with $\Lambda R<1$. The eigenvalue distribution expands as $\lambda $ is increased
($\lambda = 0.1,\ 1,\ 5$). 
b)  Cases with $\Lambda R>1$. Eigenvalue density for $\lambda =-100,\ -50, -20$.
 At large $\Lambda R$ (i.e. when $\lambda $ approaches $0^-$), 
the eigenvalues expand to  a region whose size grows linearly with $\Lambda R$.}
 \label{eigdens}
 \end{figure}

At large radius the running coupling becomes small and negative. In this regime the eigenvalues are strongly repelled from the region around zero. The effective potential for an individual eigenvalue looks like the lower curve in fig.~\ref{seff}.
We can imagine two possibilities in this case. If the potential wells near the minima are sufficiently deep, the eigenvalue distribution will consist of two disconnected patches. Clearly, a two-cut solution cannot be continuously connected to the one-cut solution at weak coupling. In this case there will be a large-$N$ phase transition at some critical coupling. We found, however, (numerically) that this scenario is not realized in $\mathcal{N}=2$ SYM. The effective potential is always sufficiently shallow, such that the short-distance repulsion between the eigenvalues spreads them along a single connected interval $(-\mu ,\mu )$ for any $\lambda$. For  a negative $\lambda $ above some critical value, the eigenvalue density gets a double-hump structure with a dip in the middle, shown in fig.~\ref{densityprofiles}.

The extent of the eigenvalue distribution grows rapidly as $\lambda \rightarrow -\infty $ (exponentially in $\lambda $ or linearly in $R$).
The series expansion (\ref{kara}) is useless in this regime, because its radius of convergence is equal to one.
In appendix A we illustrate this point by considering a toy model
obtained by truncating the series  (\ref{kara}).
The model exhibits a phase transition at a finite $R$ and, as a result, the
resulting large radius behavior is markedly different from the behavior in the actual eigenvalue density in ${\cal N}=2$ theory.

We can solve the saddle-point equation (\ref{uno}) asymptotically in  the limit $\lambda \rightarrow 0^- $, i.e. $R\rightarrow \infty $, where $\mu \rightarrow \infty $. The density will then take a scaling form: $\rho (x)= \hat{\rho }(x/\mu )/\mu $, everywhere apart from tiny regions near the endpoints. Using the asymptotic form of the kernel:
\begin{equation}
 K(x)=2x\ln \left(|x|\,{\rm e}\,^{\gamma }\right)+O\left(\frac{1}{x}\right)\ ,
\end{equation}
we can see that in the scaling limit the Hilbert kernel scales away, and we are left with the equation
\begin{equation}
 \int_{-\mu }^{\mu }dy\,\rho (y)(x-y)\ln \big( |x-y| \,{\rm e}\,^{\gamma }\big)=x\ln\Lambda R\ .
\end{equation}
Differentiating this equation twice, we then get:
\begin{equation}
 \strokedint_{-\mu }^{\mu }\frac{dy\,\rho (y )}{x-y}=0\ .
\end{equation}
This equation has a unique normalizable solution:
\begin{equation}
\label{asymrho}
 \rho (x)=\frac{1}{\pi \sqrt{\mu ^2-x^2}}\, .
\end{equation}
This can be seen to agree with the eigenvalue density \cite{Douglas:1995nw,Ferrari:2001mg} obtained from  Seiberg-Witten theory  directly in flat space.

\begin{figure}[h!]
 \centering
 \includegraphics[width=.7\textwidth]{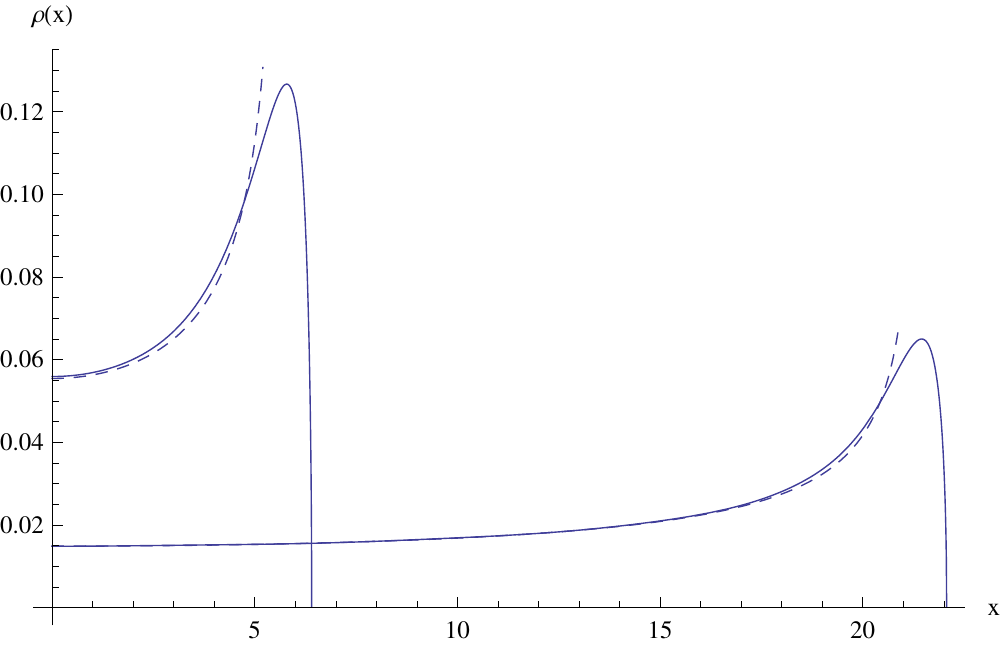} 
\caption{ \label{verylargelambda}
Eigenvalue density for $\lambda =-15,\ -10$ (solid lines, from top to bottom)
and comparison with the curves obtained from the analytic formula (\ref{asymrho}) (dashed lines).}
 \label{negative}
\end{figure}

As we will see  from the numerics, shown in fig.~\ref{verylargelambda}, the analytic solution represents a very good approximation to the large radius eigenvalue density, apart from a  small region near the endpoints, where it violates the boundary conditions. 

Returning  to the original equation, one can calculate that
\begin{equation}
 \int_{-\mu }^{\mu }dy\,\,\frac{(x-y)\ln|x-y|}{\pi \sqrt{\mu ^2-y^2}}=x\ln\frac{\mu {\rm e}}{2}\,.
\end{equation}
Consequently,
\begin{equation}
\label{yy00}
 \mu = c_0  \ {\rm e}\,^{-\frac{4\pi ^2}{\lambda }}\ 
 =c_0\ \Lambda R  \ ,\qquad c_0\equiv 2{\rm e}\,^{-1-\gamma}\ .
\end{equation}

Let us now compute the expectation value of the circular Wilson loop and the 
free energy in this large radius regime.
Using the asymptotic analytic formula (\ref{asymrho}) and (\ref{wils}), we obtain
\be
W(C_{\rm circle}) = \pi I_0(2 \pi \mu )  \approx\frac{\,{\rm const}\,}{\sqrt{\mu}} \,\,{\rm e}\,^{2\pi \mu }
\approx  \,{\rm const}\,\cdot \exp \left(4\pi \,{\rm e}\,^{-1-\gamma} \Lambda R - {1\over 2} \ln \Lambda R \right)\ ,\qquad \Lambda R\gg 1\ .
\ee
As in the $SU(2)$ case, we obtain a perimeter law.
The leading behavior can be interpreted as negative mass renormalization
$\Delta m=-2 \,{\rm e}\,^{-1-\gamma}\Lambda $ for a heavy probe particle going around the circular loop.
In terms of the running coupling constant, this is
\be
\ln W(C_{\rm circle}) \approx  4\pi \,{\rm e}\,^{-\frac{4\pi^2}{ \lambda}-1-\gamma} +\frac{2\pi^2}{\lambda} +{\rm const},\qquad \lambda \to 0^-.
\ee

To compute the free energy, we start with the relation
\begin{equation}
 \frac{\partial F}{\partial \ln R}=-2N^2\left\langle a^2\right\rangle.
\end{equation}
At large $R$, using the asymptotic density,
\begin{equation}
  \frac{\partial F}{\partial \ln R}=-N^2\mu ^2=-c_0^2\ 
  N^2\Lambda ^2R^2\ .
\end{equation}
Therefore
\begin{equation}
 F=-\frac{c_0^2}{2}\ N^2\Lambda ^2R^2\ ,
\end{equation}
at $\Lambda R\gg 1$.

%%%%%%%%%%%%%%%%%%

\subsubsection*{Numerical solution}

%%%%%%%%%%%%%%%%%%

We start with small values of $\lambda $, where we know the detailed analytic form of the solution.  We consider the ansatz
\be
\rho(x)=\left(\sum_{k=0}^{n} a_k x^{2k} \right) \sqrt{\mu^2-x^2}\ .
\ee
By taking $n$ sufficiently large, one can achieve any desired accuracy, as long as the solution contains only one cut.
We have computed the different coefficients $a_k$ and $\mu$ numerically for $n=1,2,3,...$ , and checked that the numerical solution is stable under
further increasing $n$.

\begin{figure}[h!]
\centering
\includegraphics[width=.45\textwidth]{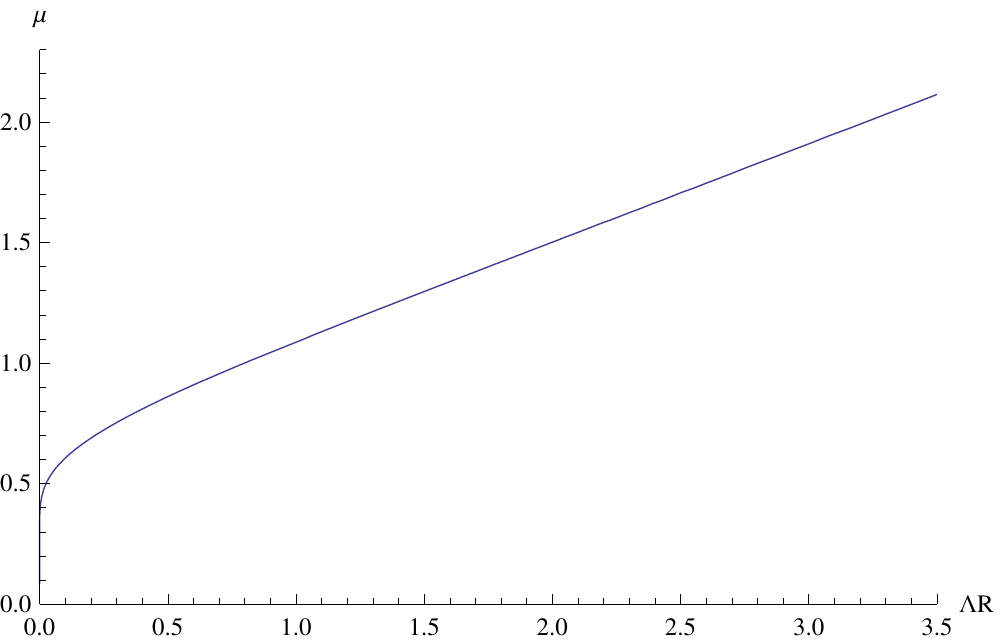}
\includegraphics[width=.45\textwidth]{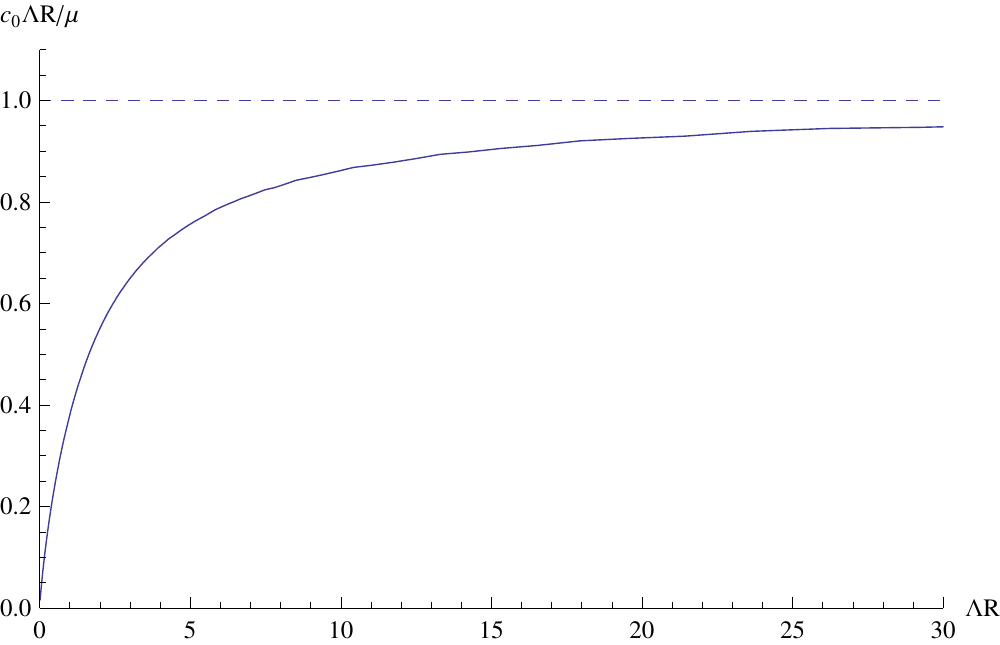} 
\caption{a) $\mu$ as a function of $\Lambda R$.
b) Comparison between the analytic asymptotic expression (\ref{yy00}) (dashed line)
and the numerical value of $\mu $ (solid line), where $c_0\equiv 2\,{\rm e}\,^{-1-\gamma} $.}
\label{mumumu}
\end{figure}

Consider first the interval $0<\lambda<\infty$, corresponding to $0<\Lambda R<1$.
Fig. \ref{eigdens}a  shows the eigenvalue density for $\lambda=0.1,\ 1,\ 5$. 
The eigenvalue distribution expands as $\lambda $ is increased.
The same qualitative shape 
 is maintained by further increasing $\lambda $ up to
$\lambda =\infty$, where $\mu $ has a finite value $\mu_\infty\sim 1.09$. 
As the equation (\ref{uno}) has an analytic dependence on the parameter $\ln \Lambda R$, there is no discontinuity
in crossing from $\Lambda R<1$ to the region $\Lambda R>1$.
Thus, at $\lambda \to -\infty$, the eigenvalue density coincides with  $\lambda \to \infty$.
The interesting behavior arises for larger values of $\lambda $.
The repulsive force $-x\ln \Lambda R$ becomes stronger and the eigenvalues are repelled from the region of small $x$.
This can be seen in fig. \ref{eigdens}b, which  shows the eigenvalue density for $\lambda=-100,\ -50,\ -20$.

The numerical results show that, unlike the toy model of appendix A, the one-cut solution exists for all values of $\lambda $. There is no phase transition.
When $\lambda /4\pi^2 $ is negative and approaches to $0^-$,
in the present case $\mu$ begins to increase exponentially with $-4\pi^2/\lambda $ (therefore linearly with $\Lambda R$);  the eigenvalue distribution gets stretched out, up to a value $\mu\gg 1$,
with an eigenvalue  density that, as $\lambda \to 0^-$, approaches the asymptotic form (\ref{asymrho}).
This is shown in fig. \ref{negative}, which displays the eigenvalue density for $\lambda = -15,\ -10$ along with the analytic asymptotic eigenvalue density  (\ref{asymrho}).
We see that there is a good agreement between the numerical and analytic curves away from the boundary $x=\pm \mu$,
an agreement which is clearly better for  $\lambda=-10$.

\begin{figure}[h!]
\centering
\includegraphics[width=.45\textwidth]{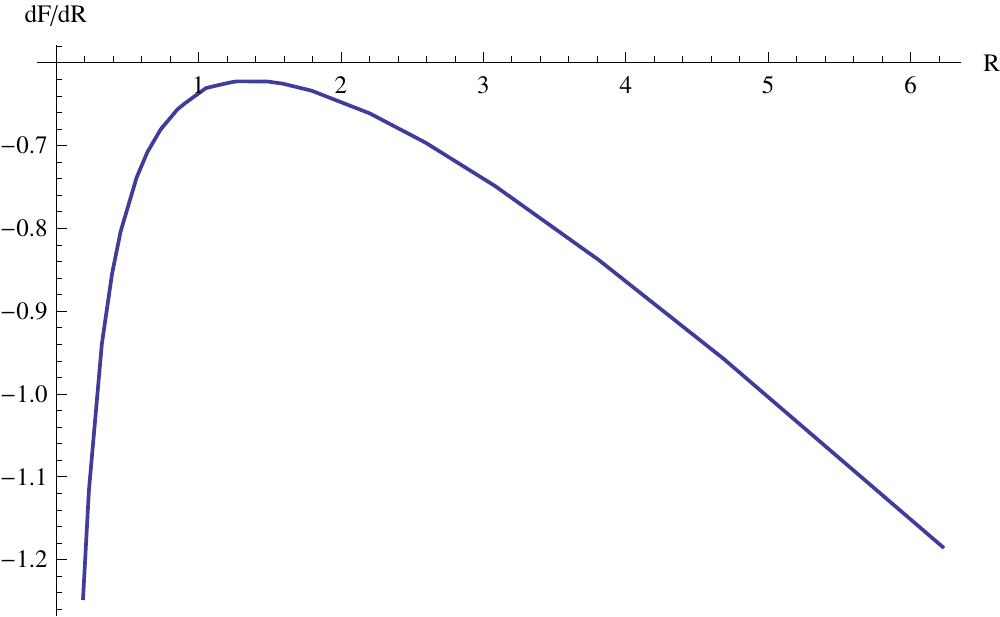}
\includegraphics[width=.45\textwidth]{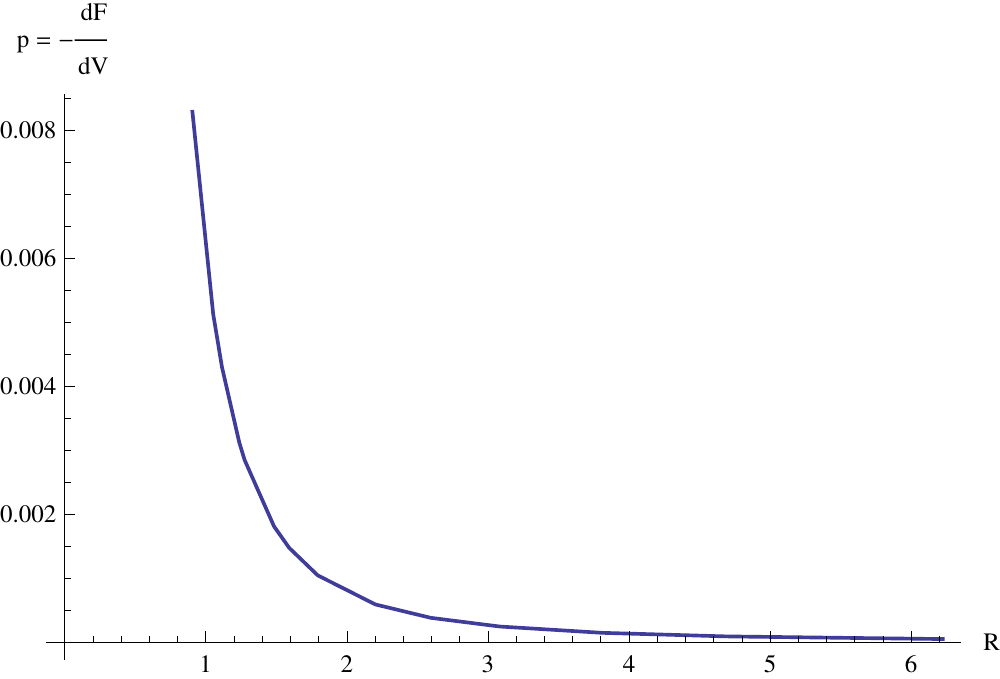} 
\caption{a) $dF/dR$ as function of $\Lambda R$.
b) The pressure, $p=-dF/dV$, as a function of the volume $V=8\pi^2R^4/3$.}
\label{freeg}
\end{figure}

Fig. \ref{mumumu}a shows $\mu$ as a function of $\Lambda R$.
This figure illustrates in particular that the point $\Lambda R=1$ is not special, as $\mu(\Lambda R)$ is smooth at this point.
It also exhibits  the linear law $\mu \sim \Lambda R$ at large $R$, obtained analytically in the previous subsection.
Fig.  \ref{mumumu}b shows that the slope of the curve asymptotically approaches the constant $c_0=2\,{\rm e}\,^{-1-\gamma} $ found earlier.
Note that the linear behavior of $\mu $ is a good approximation even in the region $\Lambda R\sim 1$. 

Finally, fig. \ref{freeg}a shows the derivative of the free energy as a function of $\Lambda R$. At $\Lambda R \gg 1 $ it exhibits the linear behavior found analytically
in the previous subsection. At $\Lambda R\ll 1$, it has the behavior $dF/dR\sim N^2/(2R\ln R)$,
as it can be easily deduced from the perturbative formulas.
We notice that there is a change of sign in $d^2F/dR^2$ at the critical radius where $d^2F/dR^2=0$.
A natural question is whether this has any thermodynamic implication.
The pressure of the system is $p= -dF/dV$, where $V$ is the volume 
of $S^4$, $V\equiv 8\pi^2R^4/3$.
In stable systems, the pressure must be positive and, in most common systems, also the compressibility,  $\kappa =-\frac{1}{V} \frac{dV}{dp}$, is positive.
This is also the case in the present system, as shown in fig. \ref{freeg}b:
the pressure has positive sign in the whole interval $0<\Lambda R<\infty$. 
The same plot shows that the compressibility
is always positive.

In conclusion, we have seen that the one-cut solution exists for all values of $\Lambda R$.
We have also checked that there are no two-cut solutions at any $\Lambda R$ (see also discussion at the end of appendix A).

\section{Remarks on superconformal theories}

The simplest $\mathcal{N}=2$ superconformal theory (SCYM) is $SU(N)$ SYM with $N_f=2N$ hypermultiplets in the fundamental respresentation. The large-$N$ limit of SCYM on $S^4$ was studied in \cite{Rey:2010ry,Passerini:2011fe,Bourgine:2011ie,Fraser:2011qa}, with the idea to get an insight into its putative AdS dual. One may expect that in the strong coupling limit it should behave similarly to $\mathcal{N}=4$ SYM. Quite surprisingly, the strong-coupling limit of  the $\mathcal{N}=2$ theory is very different from $\mathcal{N}=4$. The free energy  tends to a constant rather than growing logarithmically \cite{Bourgine:2011ie}, while the Wilson loop VEV grows as a cube of the 't~Hooft coupling \cite{Passerini:2011fe}, in a stark contrast to the exponential growth in the $\mathcal{N}=4$ SYM.

 Below we will comment on the computation of the free energy in $\mathcal{N}=2$ SCYM, in $\mathcal{N}=4$ SYM and in its holographic dual. The holographic computation is a classic result in AdS/CFT \cite{Emparan:1999pm}. Yet, we were unable to find a holographic derivation  of the $\ln \lambda $ dependence  in the literature, and so we will derive this result by reinterpreting the formulas in \cite{Emparan:1999pm}. We will argue that the logarithm arises due to a subtle difference between field-theory and supergravity UV cutoffs.

\subsection{$\mathcal{N}=4$ SYM}

The $\mathcal{N}=4$ SYM localizes to the Gaussian matrix model, whose exact eigenvalue density is given by Wigner semi-circle law:
\begin{equation}
 \rho (x)=\frac{4 }{\lambda }\,\sqrt{{\lambda }-4\pi ^2x^2}.
\end{equation}
The free energy then is
\begin{equation}\label{N=4free}
  \frac{\partial F}{\partial \lambda }=-\frac{8\pi ^2N^2}{\lambda ^2}\,\left\langle a^2\right\rangle=-\frac{N^2}{2\lambda }
  \qquad \Longrightarrow\qquad F=-\frac{N^2}{2}\,\ln\lambda ,
\end{equation}
a result which is easy to obtain directly from the Gaussian matrix integral.

\medskip

$\partial F/\partial \lambda$ represents the integrated one-point function of the action density
(more generally, $\partial^nF/\partial \lambda^n$  is related to the integrated $n$-point function of the action density).
Since $\partial F/\partial \lambda = - Z^{-1} \partial Z/\partial \lambda$, the  result is independent of the normalization
of $Z$, as long as there is no extra $\lambda $-dependent normalization factor.
This requirement is equivalent to saying that the measure does not depend on $\lambda $.

\subsection{Holographic calculation}

At strong coupling, the free energy of $\mathcal{N}=4$ SYM is identified with the on-shell supergravity action on $AdS_5$:
\begin{equation}
 I=-\frac{1}{16\pi G_N}\int_{AdS_5}^{}d^5x\,\sqrt{g}\left(R+\frac{12}{L^2}\right)+I_{\rm boundary},
\end{equation}
where $G_N$ is the 5d Newton's constant, $L$ is the radius of the Anti-de-Sitter space, and $I_{\rm boundary}$ contains the Gibbons-Hawking surface term and boundary counterterms, necessary to eliminate the divergences. 

The metric of $AdS_5$, foliated such that the boundary is $S^4$, assumes the following form:
\begin{equation}
 ds^2=\frac{L^2dr^2}{L^2+r^2}+r^2d\Omega _{S^4}^2.
\end{equation}
The boundary is the large four-sphere at $r\rightarrow \infty $. Using the Einstein's equations, $R=-20/L^2$, the action is calculated to be \cite{Emparan:1999pm}:
\begin{equation}
 I=\frac{4\pi }{3G_NL}\int_{0}^{r_0}\frac{dr\,r^4}{\sqrt{L^2+r^2}}+{\rm counterterms},
\end{equation}
where $r_0$ is the supergravity cutoff. The counterterms cancel off quartic, quadratic and logarithmic divergences in this expression. The explicit expressions can be found in \cite{Emparan:1999pm,Marino:2011nm}. 

Let us concentrate on the logarithmically divergent term: 
\begin{equation}\label{SUGRAfreeI}
 I_{\rm log}=\frac{\pi L^3}{2G_N}\,\ln\frac{r_0}{L}.
\end{equation}
The  counterterm that eliminates this log-divergence is also responsible for the trace anomaly in the energy-momentum tensor \cite{Henningson:1998gx,Balasubramanian:1999re}, so the coefficient in front  is directly related to the $a=c$ conformal anomaly of $\mathcal{N}=4$ SYM. Once this coefficient is expressed in terms of $N$ of the $SU(N)$ according to the AdS/CFT dictionary:
\begin{equation}\label{Newton's}
 \frac{\pi L^3}{2G_N}=N^2,
\end{equation}
it precisely matches with a logarithmic divergence that is also present in the partition function of $\mathcal{N}=4$ on $S^4$ \cite{Burgess:1999vb}.

The common lore of the AdS/CFT duality is that one should first eliminate  divergences  by adding counterterms \cite{Skenderis:2002wp} and then compare  finite quantities. For the free energy one would then obtain a  $\lambda $-independent constant, in a clear contradiction with the field-theory calculation. We are going to argue that the reason for this mismatch is a (finite and calculable) difference between the field-theory and supergravity counterterms, that follows from the energy-radius relation for the AdS metric \cite{Peet:1998wn}.

\begin{figure}[t]
\begin{center}
 \centerline{\includegraphics[width=6cm]{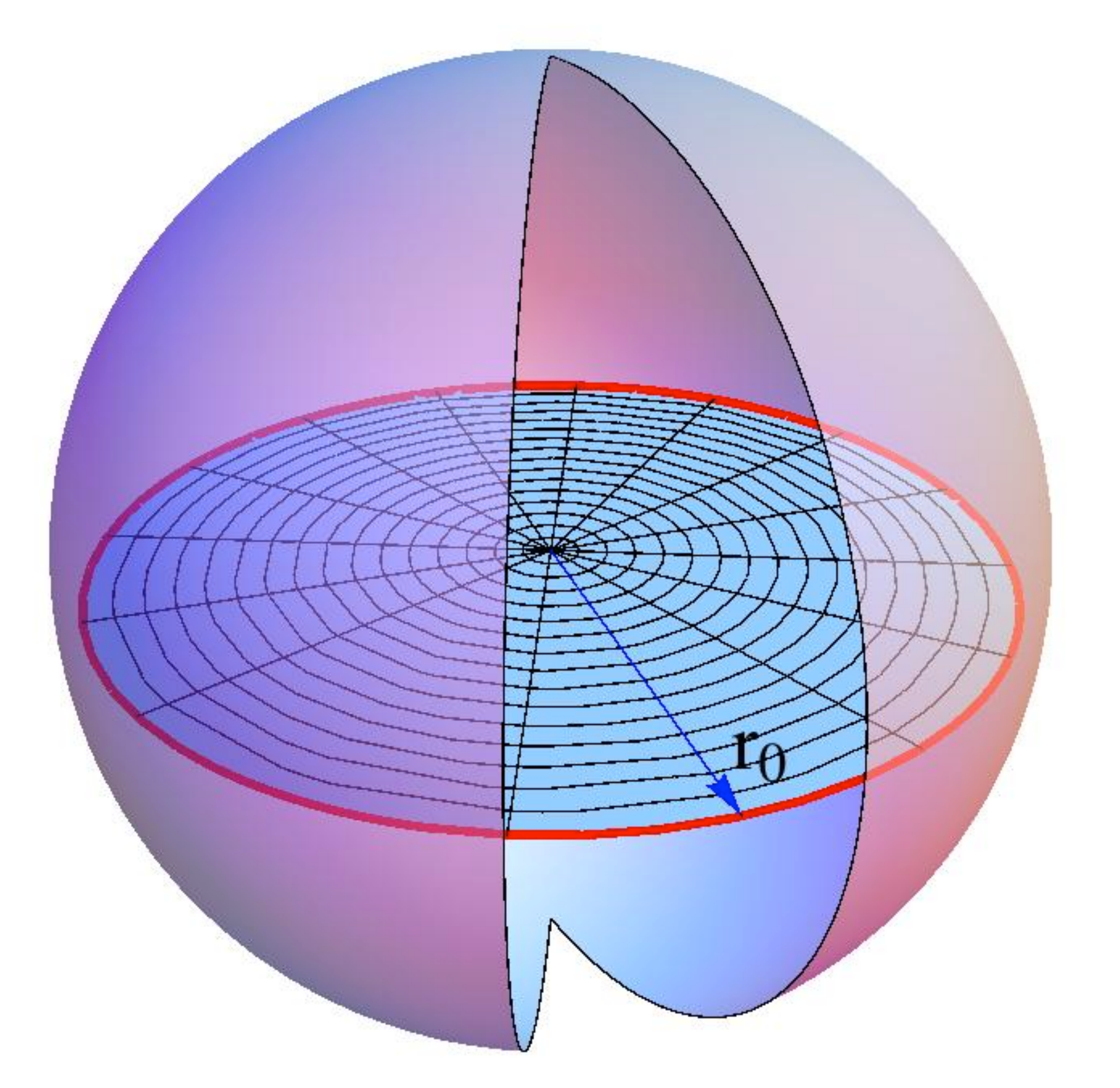}}
\caption{\label{BigCircle}\small The string worldsheet for the probe W-boson moving along a big circle of $S^4$.}
\end{center}
\end{figure}

We will relate the supergravity cutoff with the field-theory cutoff by a simple physics argument \cite{Peet:1998wn}. Consider an $SU(N+1)$ theory broken down to $SU(N)\times U(1)$ by pulling one brane out of the stack that creates AdS and placing it at some radial distance away from the horizon. The W-boson, associated with broken symmetry, is described by a string stretching from the horizon to the brane. Consider now the W-boson with the largest possible mass. The upper bounds on the mass, both in field theory and in supergravity, arise from the UV regularization. In field theory, the largest possible mass is obviously $\Lambda _0$. We can express $r_0$ through $\Lambda _0$ by computing what is the largest possible mass in supergravity. In order to do that we
consider a heavy probe with the largest attainable mass moving along the big circle of $S^4$. On the field theory side, the action of the heavy probe is $2\pi R\Lambda_0 $. The string action (the area of the worldsheet stretching from  the horizon to the cutoff surface along the big circle, fig.~\ref{BigCircle}) is
\begin{equation}
 \frac{L}{2\pi \alpha '}\int_{0}^{r_0}\frac{dr\,2\pi r}{\sqrt{L^2+r^2}}
 =\frac{Lr_0}{\alpha '}+{\rm finite},
\end{equation}
from which we conclude that
\begin{equation}\label{AdScutoff}
 2\pi R\Lambda _0=\frac{Lr_0}{\alpha '}=\frac{\sqrt{\lambda }\,r_0}{ L}\,,
\end{equation}
where in the last equality we used the standard AdS/CFT relationship between the radius of $AdS_5$ and the 't~Hooft coupling of $\mathcal{N}=4$ SYM. The necessity to introduce $\sqrt{\lambda } $ in the energy-radius relation can be also seen from the (constant) Weyl rescaling that brings the near-horizon metric of the D3 brance into the canonical $AdS_5\times S^5$ form \cite{Bianchi:2001de}.

Combining (\ref{SUGRAfreeI}), (\ref{Newton's}) and (\ref{AdScutoff}), we thus express the free energy in terms of the  field-theory quantities:
\begin{equation}\label{FSUGRA}
 F_{\rm SUGRA}=-\frac{N^2}{2}\,\ln\lambda +N^2\ln 2\pi R\Lambda _0.
\end{equation}
The second, infinite piece is renormalized away upon deriving the matrix model by localization\footnote{We were informed by V.~Pestun that the dependence on $R$, and consequently on $\Lambda _0$ can be easily reconstructed for a generic $\mathcal{N}=2$ theory, with the coefficient equal to the $a$-type anomaly. In particular, in $\mathcal{N}=4$ SYM, the $R$-dependence is consistent with (\ref{FSUGRA}).}. The finite piece  reproduces (\ref{N=4free}). 

\subsection{$\mathcal{N}=2$ SCYM}\label{sec:SCYM}

Consider now the superconformal $\mathcal{N}=2$ $SU(N)$ theory with $N_f=2N$ fundamental massless flavors.
The partition function  is
 \begin{equation}\label{matrixSCYM}
 Z= \int d^{N-1}a \,\prod_{i<j}^{}\left[\left(a_i-a_j\right)^2H^2(a_i-a_j)\right]\,{\rm e}\,^{-N\sum\limits_{i}\left(\frac { 8\pi^2 } {\lambda }\, a_i ^2 +2\ln H(a_i)\right)},
\end{equation}
which leads to the following saddle-point equations:
\begin{equation}\label{integralequationinSCYM}
 \strokedint_{-\mu }^{\mu } dy\,\rho (y)\left(\frac{1}{x-y}-K(x-y)\right)=\frac{8\pi ^2}{\lambda }\,x-K(x),
\end{equation}
The asymptotic solution at infinite coupling can be found by Fourier transform \cite{Passerini:2011fe}:
\begin{equation}\label{N2SCYMdens}
 \rho _\infty (x)=\frac{1}{2\cosh\frac{\pi x}{2}}\,.
\end{equation}

The free energy  at strong coupling  follows then from $\left\langle a^2\right\rangle_\infty =1$:
\begin{equation}
 F=\,{\rm const}\,+\frac{8\pi ^2N^2}{\lambda }+\ldots 
\end{equation}
The next correction, as argued in \cite{Bourgine:2011ie}, is proportional to $\ln^2\lambda /\lambda ^2$. We will compute the coefficient of proportionality  and also the next term in the expansion.

 The asymptotic density extends to infinity, but at any finite $\lambda $ the density is supported on a finite interval $(-\mu ,\mu )$, where
$\mu $ is determined by the equation  \cite{Passerini:2011fe}:
\begin{equation}\label{muClambda}
 C\sqrt{\mu }\,{\rm e}\,^{\frac{\pi \mu }{2}}=\lambda ,
\end{equation}
where\footnote{An analytic expression for the constant $C$ in terms of an infinite sum is given in
\cite{Passerini:2011fe}.} $C=14.60...$

The deviation of the density from its asymptotic form leads to corrections to $\left\langle a^2\right\rangle$, which are calculated in appendix~\ref{detailsSCYM}:
\begin{equation}\label{meansquare}
 \left\langle a^2\right\rangle
 =1-\frac{4\pi ^2\mu ^2}{\lambda }+\frac{b\mu ^{\frac{3}{2}}}{\lambda }
 +O\left(\frac{\mu }{\lambda }\right),
\end{equation}
where $b=0.4018...$
The origin of these two terms is easy to understand. The asymptotic density, which extends all the way to infinity, underestimates the number of small eigenvalues, the true eigenvalue distribution is more compressed. Hence the negative sign of the first correction. However, in addition to an overall increase in the density between $-\mu $ and $\mu $, which is  the leading effect, there is a short-range pile up of  the eigenvalues near the endpoints, which produces the next term with the positive sign.

Integrating the free energy with the help of (\ref{muClambda}), which, at this order, amounts to setting $\mu \approx (2/\pi)\ln\lambda $, we get the final result:
\begin{equation}
 F=\,{\rm const}\,+N^2\left[
 \frac{8\pi ^2}{\lambda }-\left({64\pi ^2\,\ln^2\lambda }
 -{2^\frac{7}{2}\pi ^{\frac{1}{2}}b\,\ln^\frac{3}{2}\lambda }
 +O\left({\ln\lambda }\right)\right)\frac{1}{\lambda ^2}
 \right]+\ldots 
\end{equation}
This is obviously very different from the simple $\ln (\lambda )$ behaviour in $\mathcal{N}=4$, which is exact and is valid at any coupling. In the SCYM, the free energy is a  non-trivial function of $\lambda $, and behaves logarithmically only in the weak-coupling limit.

%%%%%%%%%%%%%%%%%%%%%%%%%%%%%%%%%%%%%%
\section{Remarks on $\mathcal{N}=2^*$ theory}\label{2star}
%%%%%%%%%%%%%%%%%%%%%%%%%%%%%%%%%%%%%%

This theory is obtained by deformation of ${\cal N}=4$ SYM by adding a mass\footnote{Here the radius of the sphere is absorbed into the definition of the  mass of the hypermultiplet: $M\rightarrow MR$.} $M$
term for the adjoint hypermultiplet.
In this case the equation for the eigenvalues is
\be
\label{nnstar}
\dashint_{-\mu}^\mu dy \rho(y) \left(\frac{1}{x-y} -K(x-y)+{1\over 2}(K(x-y+M)+K(x-y-M))\right)= \frac{8\pi^2}{\lambda_0}\ x\ .
\ee
Now the eigenvalue density depends on a two-parameter space $\lambda_0, M$.
Some corners of this two-parameter space were studied in \cite{Russo:2012kj}.
For small $M$, it approaches the eigenvalue density of ${\cal N}=4$ SYM 
plus $O(M^2)$ corrections,
\be
\rho(x)\approx\sqrt{\mu^2-x^2}\ \frac{8\pi}{\lambda_0} (1-M^2)\ ,\qquad 
\mu=\frac{\sqrt{\lambda_0}}{2\pi\sqrt{1-M^2}}\ ,
\ee
for $M\ll 1$ and $ \lambda_0 \gg M^2$.
In particular, one finds $W(C_{\rm circle})\approx \exp \big[\sqrt{\lambda_0}\ (1+M^2/2)\big]$
and
\be
F\approx -\frac{N^2}{2} (1+M^2)\ln\lambda_0\  ,\qquad M\ll 1\ ,\ \ \ \lambda_0 \gg M^2\ .
\ee

\medskip

{}For large $M$ and fixed $M\ e^{-\frac{4\pi^2}{\lambda_0}+1+\gamma}\equiv \Lambda $, the theory flows to 
the ${\cal N}=2$  SYM theory studied in  section 2, with renormalized coupling $\lambda$ defined by $ \Lambda R \equiv  e^{-{4\pi^2\over\lambda}}$.
To show this explicitly, we take the large $M$ limit, where one can approximate
\be
2K(x-y)-K(x-y+M)-K(x-y-M)= 2K(x-y) -2K'(M)\, (x-y) +O(1/M^2)
\ee
Thus, neglecting  $O(1/M^2)$ contributions, we have
\be
\dashint_{-\mu}^\mu dy \rho(y) \left(\frac{1}{x-y} -K(x-y)+K'(M)(x-y)\right)= \frac{8\pi^2}{\lambda_0}\ x\ .
\ee
Using
\be
\dashint_{-\mu}^\mu dy \rho(y) K'(M)(x-y) =x  K'(M)\ ,
\ee
and $K'(M)\approx 2\ln Me^{1+\gamma} +O(1/M^2)$, we are left 
 with the  saddle-point equation (\ref{uno}) for the ${\cal N}=2$ pure SYM theory.

\medskip

Returning to the ${\cal N}=2^*$ theory, 
by solving the saddle-point equation  (\ref{nnstar}) 
numerically one can also explore other regions
of the two-parameter space $\lambda_0,\ M$, away from the corners represented by
the ${\cal N}=4$ and ${\cal N}=2$ SYM theories.
One novel feature with respect to the ${\cal N}=2$ theory studied in  section 2 is  that, for a wide range of parameters,
eigenvalues accumulate near $x=0$ and near $x=\pm \mu $ at the same time.

\begin{figure}[h!]
 \centering
 \includegraphics[width=.6\textwidth]{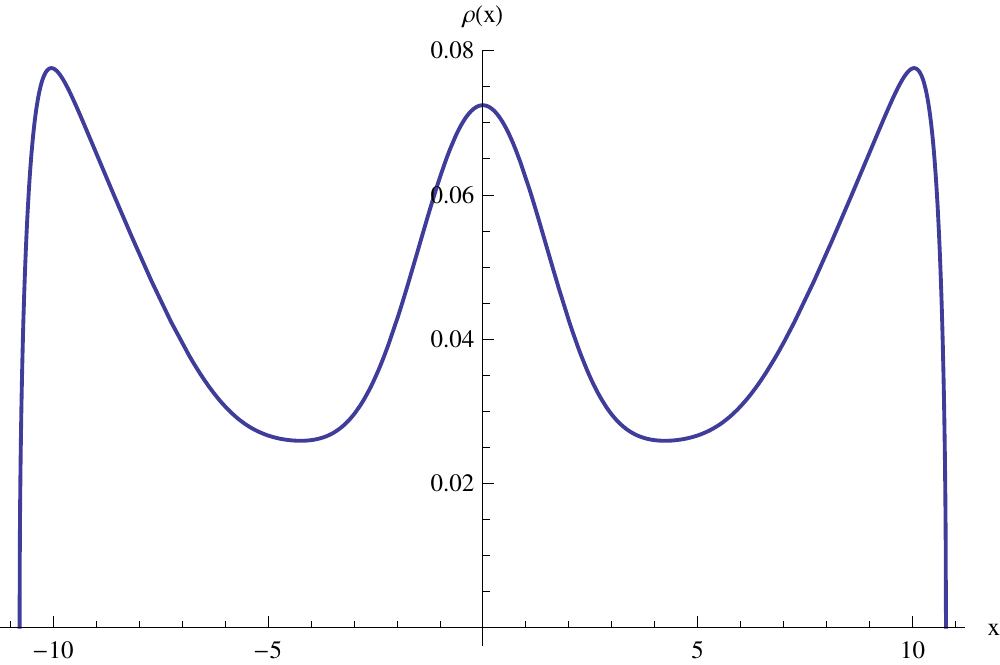} 
\caption{ Eigenvalue density for $\lambda_0=10\pi^2$, $M=9$ in ${\cal N}=2^*$ SYM theory.}
 \label{nnstarf}
\end{figure}

This is illustrated by fig. \ref{nnstarf}.
It exhibits a hill at 
$x=0$ and two more at $x=\pm \mu$. As $M$ is further increased, the hill at $x=0$ diminishes, and $\mu $ increases, approaching the eigenvalue density of ${\cal N}=2$ theory studied in the previous section.

We have also checked that  in ${\cal N}=2^*$ SYM theory the one-cut solution exists in all regions of the parameter space
and that there are no two-cut solutions in any region of the parameter space.

\section{Instanton contributions}

Instanton contributions are usually believed to be negligible at large $N$, because of the exponential factor $\exp(-8\pi ^2N/\lambda )$ in the instanton weight.
It is possible, however, that the instanton moduli integration overcomes this exponential suppression making the instanton contribution formally divergent in the large-$N$ limit. This phenomenon leads to an instanton-induced phase transition \cite{Gross:1994mr}. It was shown in \cite{Passerini:2011fe} that such a transition does not happen in the superconformal $\mathcal{N}=2$ theory discussed in sec.~\ref{sec:SCYM} and that the instantons always remain exponentially suppressed in $\mathcal{N}=2$ SCYM. Here we analyze the instanton weight in the pure $\mathcal{N}=2$ SYM.

The expression for the instanton factor in pure 
 $\mathcal{N}=2$ SYM theory can be obtained by taking a suitable limit of the instanton weight in $\mathcal{N}=2^*$ theory.
When regarding the $\mathcal{N}=2^*$ theory as a UV regularization of the pure $\mathcal{N}=2$ SYM some care is necessary in picking the same renormalization scheme in all parts of the localization partition functions. By that we mean the use of the same definition for the running coupling, which is a priori ambiguous due to the choice of the subtraction point. As in sec.~\ref{2star}, we adopt the following definition:
\begin{equation}
\label{lambdaR}
 \frac{4 \pi ^2}{\lambda }=\frac{4\pi ^2}{\lambda_0 }-\ln M-1-\gamma .
\end{equation}
where $\lambda_0 =g^2N$  represents the bare coupling of $\mathcal{N}=2^*$
and as before $\lambda $ denotes the running coupling constant of the  $\mathcal{N}=2$ theory. We emphasize that once we have used this renormalization prescription in evaluating the perturbative part of the partition function, the definition of the running coupling in the instanton weight is not a matter of choice any more. We should use the same definition. In particular, the dynamically generated scale,
\begin{equation}\label{Lambda}
 \frac{4\pi ^2}{\lambda }=-\ln \Lambda R ,
\end{equation}
 is then given by
\begin{equation}
 \Lambda R=M\,{\rm e}\,^{-\frac{4\pi ^2}{\lambda _0}+1+\gamma }.
\end{equation}

The pure $\mathcal{N}=2$ SYM  without matter multiplets arises in the limit $M\rightarrow \infty $, $\lambda_0 \rightarrow 0$ with $\Lambda $ fixed. 
The one-instanton contribution to the partition function of the $\mathcal{N}=2^*$ theory can be found e.g.  from the expression given in \cite{Okuda:2010ke} (which is a particular case of the Nekrasov partition function \cite{Nekrasov:2002qd,Nekrasov:2003rj}):
\begin{equation}
 Z^{\rm inst}_1=-\,{\rm e}\,^{-\frac{8\pi ^2N}{\lambda_0 }+i\theta_0 }M^2\sum_{l=1}^{N}\prod_{j\neq l}^{}
 \frac{\left(a_l-a_j+i\right)^2-M^2}{\left(a_l-a_j\right)\left(a_l-a_j+2i\right)}\,.
\end{equation}
Here $\theta _0$ is the theta-angle of the $\mathcal{N}=2^*$ theory.
Another  possible representation is through a contour integral:
\begin{equation}\label{intrep}
  Z^{\rm inst}_1=\,{\rm e}\,^{-\frac{8\pi ^2N}{\lambda_0 }+i\theta_0 }
  \frac{2M^2}{M^2+1}\int
 \frac{dz}{2\pi }\,\,
\prod_{j=1}^{N}\frac{\left(z-a_j\right)^2-M^2}{\left(z-a_j\right)^2+1},
\end{equation}
where the contour of integration encirlces the poles at $a_j+i$ counterclockwise.

Upon taking  the limit $M\rightarrow \infty $, $\lambda_0 \rightarrow 0$ with $\Lambda $ fixed, 
the instanton weight (\ref{intrep}) becomes
\begin{equation}\label{intrepN=2}
  Z^{\rm inst}_1=\,{\rm e}\,^{-\left(\frac{8\pi ^2}{\lambda }+2+2\gamma \right)N+i\theta }
 \int
 \frac{dz}{\pi }\,\,
\prod_{j=1}^{N}\frac{1}{\left(z-a_j\right)^2+1},
\end{equation}
with the renormalized coupling given by (\ref{lambdaR}). 
Remarkably, for odd $N$, the theta-angle also gets renormalized:
\begin{equation}
 \theta =\theta_0 +\pi N.
\end{equation}
We recall that the running coupling $\lambda$ becomes negative at $\Lambda R>1$.
In such a situation, the factor ${\rm e}\,^{-\frac{8\pi ^2N}{\lambda }}$ does not suppress the instanton contribution, but, on the contrary,  enhances it!
One might naively conclude that instantons become very important at large $N$ if $\Lambda R>1$.

To clarify this important point, we need to compute $ Z^{\rm inst}_1$.
In the large-$N$ limit, the instanton weight becomes
\begin{equation}
  Z^{\rm inst}_1=\,{\rm e}\,^{-\left(\frac{8\pi ^2}{\lambda }+2+2\gamma \right)N+i\theta }
 \int
 \frac{dz}{\pi }\,\,
 \,{\rm e}\,^{-NS_{\rm inst}(z)},
\end{equation}
where the $z$ integration can now be taken along the real axis, and
\begin{equation}
  S_{\rm inst}(z)=\int_{-\mu }^{+\mu }dx\,\rho (x)\,
 \ln\left(\left(z-x\right)^2+1\right).
\end{equation}
The $z$-integral is of the saddle-point type, with the single saddle-point at $z=0$, so we find that
\begin{equation}
 Z^{\rm inst}_1\sim \,{\rm e}\,^{-NS_{\rm eff}}
\end{equation}
with
\begin{equation}\label{seffinst}
  S_{\rm eff}=-2\ln\Lambda R+2+2\gamma +\int_{-\mu }^{+\mu }dx\,\rho (x)\,
 \ln\left(x^2+1\right).
\end{equation}

The instantons are suppressed if and only if $S_{\rm eff}>0$ for any value of $\Lambda R $, which is not at all obvious as the first term becomes negative for $\Lambda R>1$. We have checked that the full action is nevertheless positive for any $\Lambda R$, by computing the action analytically in the two limiting cases of  $\Lambda R\rightarrow 0$ and $\Lambda R\rightarrow \infty  $ and numerically at intermediate $\Lambda R$'s.

At small $\Lambda R$, the integral in (\ref{seffinst}) can be estimated as $\left\langle \ln(1+x^2)\right\rangle\sim \left\langle x^2\right\rangle\sim \mu ^2\sim \lambda \sim  -1/\ln \Lambda R$.  The action thus is large and positive.  Obviously, as $\Lambda R$ grows, the action will diminish (at first logarithmically). The question is whether it crosses zero or not. To answer this question we can study the opposite limit of very large $\Lambda R$.

When $\Lambda R \rightarrow \infty $, the bare instanton weight, $\lambda\to 0^-$ and $e^{-\frac{8\pi^2N}{\lambda}}\to \infty $, which would naively imply that  the one-instanton
contribution blows up at $N\rightarrow \infty $, but we also need to take into account the last term, the matrix-model average of $\ln(x^2+1)$. Since we know the solution of the matrix model analytically at $\Lambda R\rightarrow \infty $, we can take this solution and calculate the integral. Inserting $\rho (x)$ from (\ref{asymrho}), and taking into account that $\mu \gg 1$, we get:
\bea
 \int_{-\mu }^{+\mu }dx\,\rho (x)\,
 \ln(1+x^2) &=& 2 \ln \left(\frac{1}{2} \sqrt{\mu ^2+1}+\frac{1}{2}\right)
\nonumber\\
&=& 2\ln\frac{\mu }{2}+\frac{2}{\mu }+O\left(\frac{1}{\mu }\right)\ .
\eea
Using the explicit expression for $\mu $ from (\ref{yy00}), we finally get
\begin{equation}
 S_{\rm eff}= \frac{\,{\rm e}\,^{1+\gamma }}{\Lambda R}+O\left(\frac{1}{(\Lambda R)^2}\right).
\end{equation}
The instanton action becomes very small at large $\Lambda R$, but remains  positive. 

We have checked numerically that the instanton action is a monotonic function of $\Lambda R$ and consequently never becomes smaller than zero. In conclusion, in the large $N$ limit, the one-instanton contribution is  negligible.

\section{Conclusions}

We found that Wilson loops in $\mathcal{N}=2$ SYM on a large four-sphere obey a perimeter law with a precisely calculable coefficient. The free energy scales as $R^2$. It would be really interesting to apply localization techniques to confining theories, perhaps by breaking $\mathcal{N}=2$ supersymmetry to $\mathcal{N}=1$ as in \cite{Seiberg:1994rs}. The supersymmetry breaking unfortunately ruins localization, because the superpotential term that one should add to the Lagrangian does not commute with the supercharge used in the localization procedure.

The large-$N$ superconformal $\mathcal{N}=2$ theory with $N_f=2N$ fundamental hypermultiplets turns out to be very different from $\mathcal{N}=4$ SYM at strong coupling \cite{Rey:2010ry,Passerini:2011fe,Bourgine:2011ie,Fraser:2011qa}. The dual string theory, if such exists, should thus remain strongly coupled, even when the 't~Hooft coupling is large, in contradistinction to the $\mathcal{N}=4$ case, where at large 't~Hooft coupling the dual string theory behaves semiclassically at large $\lambda $.

%%%%%%%%%%%%%%%%%%%%%%%%%%%%%%%%
\subsection*{Acknowledgments}
We would like to thank A.~Buchel and R.~Janik for discussions and V.~Pestun for correspondence.
K.Z. would like to thank the Perimeter Institute, where this work was initiated, for kind hospitality.
The work of K.Z. was supported in part by  the RFFI grant 10-02-01315, and in part
by the Ministry of Education and Science of the Russian Federation
under contract 14.740.11.0347. J.R. acknowledges support by MCYT Research
Grant No.  FPA 2010-20807 and Generalitat de Catalunya under project 2009SGR502.

%%%%%%%%%%%%%%%%%%%%%%%%%%%%%%%%%%

\appendix

\section{Toy model with phase transition}

It is instructive to consider a toy model, 
obtained by truncating the expansion of $K(x)$ to the first term. We replace
\be
K(x)\to  2\zeta(3)x^3\ .
\label{kkk}
\ee
The equation for the eigenvalue density (\ref{uno}) can then be solved exactly by the following ansatz:
\be
\rho(x) = (n_0+cx^2)\sqrt{\mu^2-x^2}\ ,
\label{onecut}
\ee
where $n_0,\ c,\ \mu$ are parameters to be determined.
The normalization condition gives
%\be
%1=\frac{\pi\mu^2}{8}(4n_0+c\mu^2)
%\ee  i.e.
\be
n_0= \frac{2}{\pi\mu^2} - \frac{c\mu^2}{4}\ .
\ee
In order to determine $\mu$ and $c$ we use the formulas
\bea
\dashint_{-\mu}^\mu dy  (n_0+cy^2)\sqrt{\mu^2-y^2}\frac{1}{ (x-y)}  &=& \pi n_0  x + \frac{\pi}{2}\ c x (2x^2- \mu^2)\ ,
\nonumber\\
\dashint_{-\mu}^\mu dy  (n_0+ c y^2)\sqrt{\mu^2-y^2} (x-y)^3 &=&
\frac{\pi}{8}\ n_0 x\mu^2(4x^2+3\mu^2) + \frac{\pi}{16}\ c x\mu^4(2x^2+3\mu^2)\ .
\nonumber
\eea
Then the saddle-point equation (\ref{uno}) becomes
\be
 n_0  + \frac{1}{2}\ c (2x^2- \mu^2)-\zeta(3) \frac{1}{4}\  \mu^2 \left( n_0(4x^2+3\mu^2) + \frac{c}{2}\ \mu^2(2x^2+3\mu^2)\right) =\frac{8\pi}{\lambda }\ .
\ee
This leads to
\be
n_0=\frac{2}{\pi  \mu ^2}-\frac{\mu ^2 \zeta
   (3)}{2 \pi }\ ,\qquad c=\frac{2 \zeta (3)}{\pi }\ ,
\ee
and $\mu^2\equiv z$ is the positive real root of the quartic equation
\be
1 - \frac{4 \pi ^2}{\lambda } \  z=  \frac{3\zeta (3)}{2} \ 
    z^2 +  \frac{3 \zeta (3)^2}{16}  z^4 \ .
\label{gao}
\ee
Using these expressions one can compute the eigenvalue density $\rho (x)$ for different
values of $\lambda $.

\smallskip 

For small $\lambda >0$, the first few terms in the perturbative expansion are as follows
\be
\mu^2 =\frac{\lambda  }{4 \pi ^2}-\frac{3
   \zeta (3) \lambda  ^3}{128 \pi ^6}+\frac{69  \zeta (3)^2\lambda  ^5}{16384 \pi ^{10}} +O(\lambda ^7)
\ee
This shows that, in this weak coupling regime, all the eigenvalues $x$ are confined in a small region $O(\sqrt{\lambda })$, so that
the term $K(x)=O(x^3)$ is small. Comparing with  (\ref{poi}), we see that in this regime the difference
between the toy model and the actual ${\cal N}=2$ theory  begins at order $O(\lambda ^4)$.

\smallskip 

When $\lambda\to \pm\infty $, $\mu $ approaches the asymptotic value,
\be
\mu^2\bigg|_{\lambda=\pm \infty} =2\sqrt{ \left(\frac{2}{\sqrt{3}}-1\right)\frac{1}{\zeta(3)}}
\cong 0.72\ ,
\label{asymu}
\ee
and
\be
n_0\bigg|_{\lambda=\pm \infty}=\frac{2}{\pi } \left(\sqrt{3}-1\right)
   \sqrt{\left(\frac{2}{\sqrt{3}}+1\right)
   \zeta (3)}\approx 0.75\ .
\ee

\smallskip 

 As $\lambda $ is increased from $-\infty $, the eigenvalues begin to accumulate
near some $x\neq 0$. The eigenvalue density at $x=0$ begins to decrease until
it vanishes at the critical value where $n_0=0$.
This occurs at
\be
\lambda_{\rm cr} =-\frac{\pi^2}{\sqrt{\zeta(3)}}\ ,\qquad \mu^2_{\rm cr}=\frac{2}{\sqrt{\zeta(3)}}\ .
\ee

\smallskip 

{}For $\lambda_{\rm cr}<\lambda <0 $, the system no longer admits the one-cut solution (\ref{onecut}).
Now the eigenvalue density  has support in two
disconnected regions. The new solution is
\be
\rho(x) = n_1 \sqrt{(a^2-x^2)(x^2-b^2)x^2}\ .
\ee
To perform the integrals in (\ref{uno}) we now use 
\bea
 \int dy \sqrt{(a^2-y^2)(y^2-b^2)y^2} &=& \frac{\pi}{8}(a^2-b^2)^2
\nonumber\\
 \int dy \sqrt{(a^2-y^2)(y^2-b^2)y^2} \frac{1}{x-y} &=& \frac{\pi}{2}\ x(2x^2-a^2-b^2)
\nonumber\\
\int dy \sqrt{(a^2 - y^2)(y^2-b^2)y^2}\ (x-y)^3 &=& \frac{\pi}{16}\ x (a^2-b^2)^2 (2x^2+3a^2+3b^2)
\eea

Solving the normalization condition and (\ref{uno}) we find
\be
a^2 = -\frac{\pi^2}{\lambda \zeta(3)} + \frac{1}{\sqrt{\zeta(3)}} \ ,\ \ \ b^2 = -\frac{\pi^2}{\lambda \zeta(3)} - \frac{1}{\sqrt{\zeta(3)} }
\ ,\ \ \ 
n_1= \frac{2 \zeta (3)}{\pi }\ .
\ee
Since $a^2$ and $b^2$ must be positive, the solution exists only for  $\lambda_{\rm cr}<\lambda <0$.
Thus this solution appears precisely at the point the one-cut solution (\ref{onecut}) ceases to exist.

Let us now compute the free energy before and after the phase transition.
The free energy is given by the formula
\be
F= \frac{8\pi^2N^2}{\lambda} \int dx \rho(x) x^2 - \frac{1}{2} N^2 \int dx\int dy
\rho(x) \rho(y) \left( \ln (x-y)^2 + \ln  H^2(x-y)\right)
\ee
In the toy model,
\be
\ln  H^2(x) \to -\zeta(3)x^4\ .
\ee

To understand the character of the phase transition, we only need the derivative with respect to $\lambda $.
On the solution, we have
\be
{dF\over d\lambda } = {\partial F\over \partial \lambda } = - \frac{8\pi^2N^2}{\lambda^2} \int dx \rho(x) x^2\ .
\ee

In the phase I lying in the regions  $0<\lambda<\infty  $ and $-\infty<\lambda< \lambda_{\rm cr}$, we have
\be
\int dx \rho_{\rm I} (x) x^2 = \frac{\mu^2}{16}(4+\zeta(3)\mu^4)\ .
\ee
where $\mu=\mu(\lambda)$ is defined by the positive real root of (\ref{gao}).
In phase II, lying in the region  $\lambda_{\rm cr}<\lambda <0$, we have
\be
\int dx \rho_{\rm II} (x) x^2 =-\frac{\pi^2}{\lambda \zeta(3)}\ .
\ee
Therefore
\be
{dF_{\rm I}\over d\lambda } =  - \frac{8\pi^2N^2}{\lambda^2} \frac{\mu^2}{16}(4+\zeta(3)\mu^4)
\ ,\qquad
{dF_{\rm II}\over d\lambda } =   \frac{8\pi^2N^2}{\lambda ^2} \frac{\pi^2}{\lambda\zeta(3)}\ .
\ee
By further differentiation, we can compute second and third derivatives of the free energy.
At the critical point $ \lambda_{\rm cr} =-\frac{\pi^2}{\sqrt{\zeta(3)}}$ we find
\bea
&& {dF_{\rm I}\over d\lambda } = {dF_{\rm II}\over d\lambda } =- \frac{8N^2 \sqrt{\zeta(3)}}{\pi^2}\ ,
\nonumber\\
&& {d^2F_{\rm I}\over d\lambda^2 } = {d^2F_{\rm II}\over d\lambda^2 } = -\frac{24 N^2\zeta(3)}{\pi^4}\ ,
\nonumber\\
&& {d^3F_{\rm I}\over d\lambda^3 } = -\frac{98N^2 \zeta(3)^{3\over 2}}{\pi^6}\ ,\qquad      {d^3F_{\rm II}\over d\lambda ^3} = -\frac{96N^2 \zeta(3)^{3\over 2}}{\pi^6}\ .
\eea
Thus there is a discontinuity in the third derivative of the free energy: the phase transition is of the third kind.

\medskip

Next, we compute the expectation value of the circular Wilson loop in both phases I and II. We use
\be
W(C_{\rm circle}) =\int dx \rho(x) \,{\rm e}\,^{2\pi x}\ .
\ee
Hence, in phase I, with eigenvalue density (\ref{onecut}), we find
\be
W_{\rm I}(C_{\rm circle}) =\frac{1}{4 \pi ^2 \mu } \left(    \pi  I_1(2 \pi  \mu ) \left(3 \mu ^4 \zeta (3)+4\right)-6 \mu ^3 \zeta (3)
   I_2(2 \pi  \mu )\right)\ .
\label{WI}
\ee
At small $\lambda $, we find the expansion
\be
W_{\rm I}(C_{\rm circle}) =1+\frac{\lambda }{8}+\frac{\lambda ^2}{192}+\lambda ^3 \left(\frac{1}{9216}-\frac{5
   \zeta (3)}{512 \pi ^4}\right)+O\left(\lambda ^{4}\right)\ .
\ee
Up to this order, there is no difference between the toy model and ${\cal N}=2$ SYM. The difference  starts at the 
$\lambda^4$ terms (cf. (\ref{gwil})).
As $\lambda \to\infty $, $W_{\rm I}(C_{\rm circle}) $ approaches a finite constant, since $\mu$ approaches the asymptotic value 
(\ref{asymu}).

\medskip

In phase II, we have
\be
W_{\rm II}(C_{\rm circle}) =\frac{2\zeta(3)}{\pi} \int dx \sqrt{(a^2-x^2)(x^2-b^2)x^2}\ e^{2\pi x}\ .
\ee
We  consider two limits: 

\smallskip

\noindent a) When $\lambda \sim \lambda_{\rm cr}$, then $b\sim 0$, the integral can be computed with the result 
\be
W_{\rm II}(C_{\rm circle}) =
\frac{a^2 \zeta (3)}{2 \pi ^2} (2 \pi  a I_1(2 a \pi )-3 I_2(2 a \pi ))
\ ,\qquad 
 a^2={2\over\sqrt{\zeta (3)}}\ .
\ee
This of course agrees with (\ref{WI}) at the critical point since at this point the eigenvalue densities $\rho_{\rm I}$ and $\rho_{\rm II}$  are the same.

\smallskip

\noindent b) When $\lambda\to 0^-$, we have that $a, b\to\infty$. By rescaling the integration variable, the integral  then collapses to a point and can be easily computed. 
We find 
\be
W_{\rm II}(C_{\rm circle}) \approx  {2\over\pi}\ 
\exp[2\pi^2/\sqrt{-\lambda\zeta(3)}] \ .
\ee
In terms of $\Lambda R$, this is 
\be
W_{\rm II}(C_{\rm circle}) \approx {2\over\pi}\ \,{\rm e}\,^{{\pi\over\sqrt{\zeta(3)}} \sqrt{\ln  (\Lambda R)}}\ .
\ee
Thus the behavior of the Wilson loop in this toy model completely deviates  from the behavior of the Wilson loop in the original SYM system of section 2,
where $W(C_{\rm circle}) $ obeyed a perimeter law.

One may also consider a more general toy model obtained by replacing $K\to K_\epsilon $ in (\ref{uno}) where
\be
K_\epsilon (x) \equiv \epsilon^{-2}\  x \left( \psi(1+i \epsilon x) +\psi(1-i \epsilon x)+2\gamma \right)
\ee
This interpolates between the toy model (\ref{kkk}), recovered by taking $\epsilon\to 0$ -- since $K_\epsilon (x)=2\zeta(3)x^3+ O(\epsilon^2)$ --, 
and the actual ${\cal N}=2$ matrix model, corresponding to $\epsilon \to 1$.
{}For sufficiently small $\epsilon $ the model still undergoes a phase transition towards a phase represented by the two-cut solution.
A natural question is for which value of $\epsilon $ the two-cut solution disappears.
One obtains the following picture.
Numerically, for sufficiently small $\epsilon$, one finds two-cut solutions for $\lambda$ in an  interval $(\lambda_{\rm cr},0^-)$, with 
a similar $\lambda_{\rm cr}$.
However, as $\epsilon $ is increased, one reaches a critical value of $\epsilon \sim 0.3$ beyond which
  the two-cut solution does not exist.
In particular, it does not exist in the case of interest, $\epsilon =1$.

\section{Free energy in $\mathcal{N}=2$ SCYM}\label{detailsSCYM}

Here we present the derivation of eq.~(\ref{meansquare}). 
The starting point is an approximate solution for the density obtained in \cite{Passerini:2011fe}:
\begin{equation}
 \rho (x)=\int_{-\infty }^{+\infty }\frac{d\omega }{2\pi }\,\,\,{\rm e}\,^{-i\omega x}\rho (\omega )\ ,
\end{equation}
with
\begin{eqnarray}\label{densityfin1}
 \rho (\omega )&=&\frac{1}{\cosh\omega }+\frac{2\sinh^2\frac{\omega }{2}}{\cosh\omega }\,F(\omega )
\nonumber \\ &&
-G(\omega )\,{\rm e}\,^{i\mu \omega }\sum_{n=1}^{\infty }\frac{r_n\,{\rm e}\,^{-\mu \nu _n}F(-i\nu _n)\left(\nu _n-\nu _0\right)\left(\omega -i\nu _0\right)}{\left(\nu _n+\nu _0\right)\left(\omega +i\nu _0\right)\left(\omega +i\nu _n\right)}\,,
\end{eqnarray}
where
\begin{equation}\label{Fomega}
 F(\omega )=\frac{8\pi ^2\mu J_1(\mu \omega )}{\lambda \omega }-\frac{2a \mu J_0(\mu \omega )}{\lambda }\,.
\end{equation}
The function
\begin{equation}\label{kerns}
 G(\omega )=\frac{\sqrt{8\pi^3 }\,2^{- i\omega /\pi }\Gamma \left(\frac{1}{2}+\frac{i\omega }{\pi }\right)}{\omega \Gamma^2 \left(\frac{i\omega }{2\pi }\right)}
\end{equation}
is analytic in the lower half-plane, and has simple poles at $\omega =i\nu _n$,
\begin{equation}\label{poles}
 \nu _n=\pi \left(n+\frac{1}{2}\right),
\end{equation}
with residues
\begin{equation}\label{residues}
 r_n\equiv \mathop{\mathrm{res}}_{\omega = i\nu _n}G(\omega )
 =\frac{\left(-2\right)^{n+1}\Gamma ^2\left(\frac{n}{2}+\frac{5}{4}\right)}{\sqrt{\pi }\left(n+\frac{1}{2}\right)\Gamma \left(n+1\right)}\,.
\end{equation}
The constant $a $ has a numerical value
 $a =1.0232...$ Its representation through a combination of infinite sums can be found in \cite{Passerini:2011fe}.

The approximate density was obtained with the help of the Wiener-Hopf method and is pretty accurate on the whole interval $(-\mu ,\mu )$, except for a small vicinity of the left endpoint, $-\mu $. In particular, it violates the boundary condition $\rho (-\mu)=0 $ and thus should be used with caution. A safe strategy is to use it only for positive $x$.

The first term in (\ref{densityfin1}) is the Fourier transform of the asymptotic density (\ref{N2SCYMdens}). The second term is the first correction to the density due to the finiteness of $\mu $. This term is small when $\mu $ is large, but  affects the density evenly on the whole interval $(-\mu ,\mu )$. The last term, on the contrary, is rapidly oscillating, and consequently is only important in a small vicinity of the endpoints. The purpose of this term is to subtract the poles in the lower half-plane, thus ensuring that $\rho (x)=0$ for $x>\mu $. 

When using the approximate solution above, all averages should be computed as integrals over positive $x$. In particular,
\begin{equation}
 \left\langle a^2\right\rangle=2\int_{0}^{\mu }dx\,\rho (x)x^2=-\int_{-\infty }^{+\infty }\frac{d\omega }{\pi i}\,\,\frac{\rho'' (\omega )}{\omega -i0}\ .
\end{equation}
For the first two terms in (\ref{densityfin1}), which are manifestly even functions of $\omega $, we can use the  Sokhotski-Plemelj formula, giving in this case the expected $\rho ''(0)$. This trick does not work with the last term, because the principal value integral does not vanish. Instead, the last term should be evaluated by closing the contour in the upper half-plane and computing the residues:
\begin{equation}
 \left\langle a^2\right\rangle = 1-F(0)+\sqrt{2}\mu \sum_{n=1}^{\infty }
 \frac{r_n\,{\rm e}\,^{-\mu \nu _n}F(-i\nu _n)\left(\nu _n-\nu _0\right)}{\nu _n\left(\nu_n+\nu_0 \right)}\,,
\end{equation}
where we have only kept the leading terms at $\mu \rightarrow \infty $. Taking into account that at large $\mu $
\begin{equation}
 \,{\rm e}\,^{-\mu \nu }F(-i\nu )
 \simeq 
 \frac{1}{\lambda }\sqrt{\frac{32\pi ^3\mu }{\nu ^3}}
 -\frac{a}{\lambda }\sqrt{\frac{2\mu }{\pi \nu }}\,,
\end{equation}
we  get the formula in the text, eq.~(\ref{meansquare}),
with
\begin{equation}
 b=\frac{2^{\frac{11}{2}}}{\pi }\sum_{n=1}^{\infty }
 \frac{r_nn}{\left(2n+1\right)^{\frac{3}{2}}\left(n+1\right)}\left[
 1-\frac{a\left(2n+1\right)}{8\pi }
 \right]=0.4018...
\end{equation}

\bibliographystyle{nb}
%\bibliography{refs}

\end{document}